\documentclass[twocolumn,showpacs,preprintnumbers,amsmath,amssymb,superscriptaddress]{revtex4}

\usepackage{graphicx}
\usepackage{epstopdf}
\usepackage{dcolumn}
\usepackage{bm}

\begin{document}

\title{Universal Phase Diagram and Reentrant Superconductivity \\
of $Bi$layer Hydrated Na$_x$CoO$_2\cdot y$H$_2$O}

\author{Hiroto Ohta, Chishiro Michioka, Yutaka Itoh, and Kazuyoshi Yoshimura}
\affiliation{Department of Chemistry, Graduate School of Science, Kyoto University, Kyoto 606-8502, Japan}

\date{\today}%

\begin{abstract}
We report duration dependence of superconductivity in a 43 \% humidity atmosphere and $^{59}$Co nuclear quadrupole resonance (NQR) of polycrystalline samples of $bi$layer hydrated Na$_x$CoO$_2 \cdot y$H$_2$O.
We found a reentrant behavior of superconductivity with respect to the time duration.
We obtained the universal phase diagram where two superconducting phases
and an in-between magnetic phase are classified by $^{59}$Co NQR frequency.
Zero field $^{59}$Co nuclear spin-lattice relaxation rates $1/T_1$ indicated a magnetic critical slowing down at 5 K in the magnetic phase, excluding a charge density wave ordering, and an enhancement at and just above $T_\mathrm{c}$ in a new superconducting phase, suggesting a coexistence of magnetic ordering.
\end{abstract}


\maketitle

$Bi$layer hydrated sodium cobaltate \mbox{Na$_x$CoO$_2\cdot y$H$_2$O} ($x \sim$ 0.35, $y \sim$ 1.3) is superconductor with an optimal superconducting transition temperature $T_\mathrm{c} =$ 4.7 K\cite{Takada_Nature}, a hexagonal crystal symmetry (space group: $P6_3/mmc$) as the same as its parent compound $\gamma$-Na$_x$CoO$_2$, and the CoO$_2$ layers largely separated by intercalated H$_2$O molecules.
The layered structure of a Co triangular lattice in the edge-sharing CoO$_6$ octahedrons provides us a good opportunity to study spin frustration effects on the itinerant-electron magnets.
However, the chemical diversity including Na$^+$ ions, H$_2$O molecules and H$_3$O$^+$ oxonium ions, which are thought to partly occupy the Na site\cite{Takada_oxonium}, prevents us from full understanding the frustration effects.

The duration dependence of the superconductivity in a humidity atmosphere indicates that the $bi$layer hydrated system is in a thermodynamically nonequilibrium state at room temperature, although the bulk and microscopic properties are changing with the duration\cite{ohta_Duration, Michioka_NQR}.

In this Letter, we report reentrant superconductivity with respect to duration in a 43 \% humidity atmosphere.
The various duration dependence of the superconductivity of $bi$layer hydrated samples with various soft chemical treatments are found to be unified in the universal phase diagram classified by $^{59}$Co nuclear quadrupole resonance (NQR) frequency at 77 K and characterized by the temperature dependence of $^{59}$Co NQR nuclear spin-lattice relaxation times $T_1$'s.

\begin{figure}[tb]
\includegraphics[width=9cm]{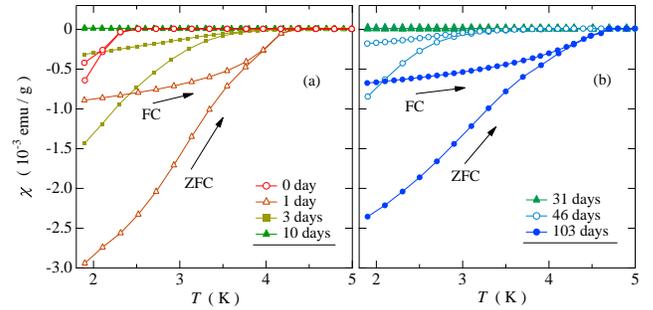}
\caption{
Temperature dependence of magnetic susceptibility $\chi$ of the A samples measured at 20 Oe after zero field cooling (ZFC) or field cooling (FC) for (a) the samples with the durations of within 10 days and for (b) the samples with the durations of over 31 days. 
Superconductivity of the A sample recovered after about one month.
}
\label{Fig:chiT_reentrantSC}
\end{figure}

\begin{figure}[tb]
\includegraphics[width=8cm]{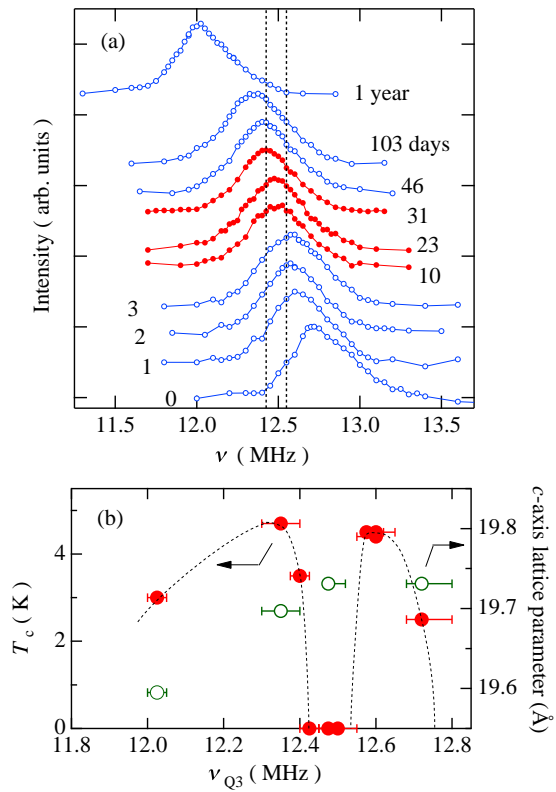}
\caption{
(a) Duration dependences of $^{59}$Co NQR ($I_z = \pm 5/2 \leftrightarrow \pm 7/2$) spectra of the A samples measured at 77 K.
(b) $\nu _\mathrm{Q3}$ (estimated at 77 K) dependence of $T_\mathrm{c}$ (closed circles) and $c$-axis lattice parameter (open circles) of the A samples.
Dotted lines are guides for the eyes.
Two superconducting phases and the in-between nonsuperconducting phase are classified by $\nu _\mathrm{Q3}$.
To be exact, the symbols at $T_\mathrm{c}$ = 0 in (b) (and Fig. \ref{Fig:Tcvsdays} and \ref{Fig:PD}) represent $T_\mathrm{c} <$ 1.9 K or nonsuperconductors.
}
\label{Fig:NQRspe}
\end{figure}

Starting polycrystalline samples of Na$_x$CoO$_2$ ($x \sim$ 0.7) were synthesized by a solid state reaction method \cite{ohta_Duration, sakurai_NCO}. 
In the previous studies, the powdered samples of Na$_{0.7}$CoO$_2$ were immersed in Br$_2$/CH$_3$CN solution to deintercalate Na$^+$ ions and then in distilled water to intercalate H$_2$O molecules.
In the present study, the powdered samples of Na$_{0.7}$CoO$_2$ were immersed in bromine aqueous solution (Br$_2aq$) of about 1.5 $\times$ 10$^{-1}$ mol/L and were stirred for 4 hrs, after Ref. \cite{Barnes}.
We consider that the simultaneous reaction of Na$^+$ deintercalation and H$_2$O intercalation is driven by Br$_2aq$, and that the different soft chemical treatment sets the different initial states of the $bi$layer hydrated samples.

We prepared two series of the $bi$layer hydrated samples labeled by A and B, which were immersed in fresh Br$_2aq$ and Br$_2aq$ aged about one year, respectively.
We expect that the degree of the intercalation reaction in the fresh Br$_2aq$ is different from that in the aged Br$_2aq$.
After filtration and 3 times wash by distilled water, the powdered samples were exposed in about 43 \% relative humidity atmosphere controlled by use of saturated aqueous solution of K$_2$CO$_3$.
Here, we expect that the lower humidity of 43 \% promotes the chemical change of the samples with the duration faster than the higher 75 \%\cite{ohta_Duration}.
Thus, two series of the $bi$layer hydrated A and B samples were systematically prepared and each duration effect on $T_\mathrm{c}$ was observed as a function of the keeping time in the humidity-controlled chamber in a daily basis, i.e., the duration of $n$ (= 0, 1, 2, $\cdots$ , 103) days.
Each sample was quickly taken out from the chamber and was frozen in a freezer ($-$18 $^{\circ}$C) to prevent further changes using a cryopreservation method.

From powder X-ray diffraction measurements, the obtained samples were confirmed as the single phase of $bi$layer hydrated Na$_x$CoO$_2\cdot y$H$_2$O.
Magnetic susceptibility $\chi$ was measured by a superconducting quantum interference device magnetometer. 
Zero field $^{59}$Co NQR spectra were measured by a spin-echo method with a phase coherent-type pulsed spectrometer. 
$^{59}$Co nuclear spin-lattice relaxation rate 1/$T_1$ was measured by an inversion recovery technique at $^{59}$Co NQR of $I_z = \pm 5/2 \leftrightarrow \pm 7/2$.

Figure \ref{Fig:chiT_reentrantSC} shows temperature dependence of magnetic susceptibilities $\chi$'s of the A samples, which were measured with temperature increasing after zero field cooling (ZFC) or field cooling (FC) at 20 Oe. 
Figure \ref{Fig:chiT_reentrantSC} (a) shows $\chi$'s of the samples with the durations of 0, 1, 3, and 10 days. 
In our previous samples in a 75 \% humidity, $T_\mathrm{c}$ increases from below 1.9 K to 4.5 K within one week and then slowly decreases down to below 1.9 K after at least one year\cite{ohta_icm}.
In the present A samples in a 43 \% humidity, $T_\mathrm{c}$ increases quickly from the initial value of 2.8 K to 4.5 K within one day and then decreases down to below 1.9 K after at least 10 days.
The different duration dependences of $T_\mathrm{c}$ demonstrate that the intercalation reaction driven by Br$_2aq$ changes the initial state of the $bi$layer hydrated samples and that the 43 \% humidity actually promotes the duration effect faster than the 75 \% humidity.

We measured $\chi$'s of the A samples for longer duration than 10 days and discovered that the A samples exhibit reentrant behaviors into a superconducting phase after about one month in 43 \% humidity atmosphere. 
Figure \ref{Fig:chiT_reentrantSC}(b) shows $\chi$'s of the A samples with the durations of 31, 46, and 103 days. 
The sample kept for 31 days did not show any superconductivity at least down to 1.9 K, whereas the sample kept for 46 days showed superconductivity with $T_\mathrm{c} \sim$ 3.5 K once again. 
After 103 days, the sample showed the maximum $T_\mathrm{c}$ = 4.7 K.

\begin{figure}[b]
\includegraphics[width=7.5cm]{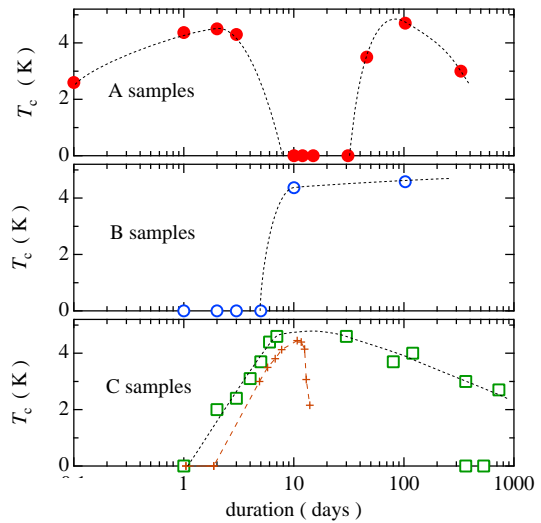}
\caption{
Duration dependences of $T_\mathrm{c}$'s of the A, B and C (previously reported\cite{ohta_Duration, ohta_icm}) samples.
Dotted and dashed lines are the guides for the eyes.
Crosses in the bottom panel are reproduced from Ref. [6]. 
}
\label{Fig:Tcvsdays}
\end{figure}

$^{59}$Co NQR frequency $\nu _\mathrm{Q}$ has been shown to characterize the $bi$layer hydrated samples more sensitively than X-ray diffraction pattern\cite{ohta_Duration, Ihara_NQR, Michioka_NQR}.
An NQR frequency reflects both local charge distribution and the deviation from cubic symmetry.
Figure \ref{Fig:NQRspe}(a) shows $^{59}$Co NQR ($I_z = \pm 5/2 \leftrightarrow \pm 7/2$) spectra of the A samples measured at 77 K.
Open and closed circles correspond to the superconducting and the nonsuperconducting samples, respectively. 
In Fig. \ref{Fig:NQRspe}(a), the samples whose $^{59}$Co NQR frequencies are located between dashed lines are nonsuperconductors.
The peak of $^{59}$Co NQR spectrum systematically shifts to a lower frequency side as the duration increases. 
This tendency has also been seen in our previous report\cite{Michioka_NQR}, although their starting NQR frequencies were different from each other.
The $bi$layer hydrated samples synthesized by using Br$_2aq$ are found to start from a higher NQR frequency than those synthesized by the original method\cite{ohta_Duration, Michioka_NQR}.

Figure \ref{Fig:NQRspe}(b) shows $\nu _\mathrm{Q3}$ dependence of $T_\mathrm{c}$ and $c$-axis lattice parameter of the A samples, where $\nu _\mathrm{Q3}$ is the peak frequency of $^{59}$Co NQR spectrum ($I_z = \pm 5/2 \leftrightarrow \pm 7/2$) estimated at 77 K.
The $T_\mathrm{c}$ and the $c$-axis lattice parameter are shown by closed and open circles, respectively.
Two superconducting phases and a nonsuperconducting phase are classified by $\nu _\mathrm{Q3}$ in this $T_\mathrm{c}$ vs $\nu _\mathrm{Q3}$ phase diagram. 
A part of this $\nu _\mathrm{Q3}$ dependence of $T_\mathrm{c}$ was observed for the accidental samples\cite{Ihara_NQR} and the systematic ones at 10 K\cite{ohta_Duration, Michioka_NQR}.
The value of the $c$-axis lattice parameter decreased with decreasing $\nu _\mathrm{Q3}$, while the $a$-axis lattice parameter remained unchanged against $\nu _\mathrm{Q3}$ (not shown).
Our result is consistent with the data reported in Ref. \cite{Ihara_Mthesis}.
This relation shows the strong correlation between thickness of the crystal along the $c$-axis and the local electric environment of Co sites.
We hereafter adopt $\nu _\mathrm{Q3}$ as the parameter to characterize the electronic state of the $bi$layer hydrated samples instead of the $c$-axis lattice parameter, since $\nu _\mathrm{Q3}$ changes in a relatively wider range compared with $c$-axis lattice parameter.
Furthermore, we are afraid of misestimation of the $c$-axis lattice parameters coming from the possible decay of samples during the powder X-ray diffraction measurement at room temperature.

Figure \ref{Fig:Tcvsdays} shows the duration dependences of $T_\mathrm{c}$'s of the A, B, and the previously reported C samples\cite{ohta_Duration, ohta_icm}.
The various duration dependences of $T_\mathrm{c}$'s are demonstrated to be produced by the controllable soft chemical treatments.

\begin{figure}[tb]
\includegraphics[width=8cm]{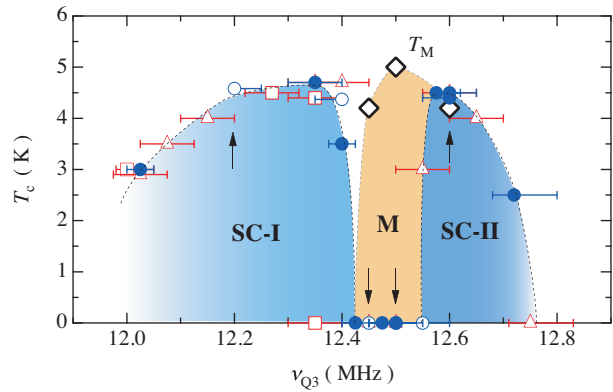}
\caption{
$T_\mathrm{c}$ vs $\nu _\mathrm{Q3}$ phase diagram at $T$ = 77 K. 
Solid and open circles and open squares correspond to the A, B and C samples, respectively.
Open triangles are the samples in the different batches synthesized by the same way as the A samples.
Two superconducting phases, labeled by SC-I and SC-II, are separated by a magnetic phase, labeled by M. 
Open diamonds are magnetic transition temperatures $T_\mathrm{M}$'s.
Dashed lines are guides for the eyes.
The samples indicated by arrows are served to measure $^{59}$Co NQR $T_1$ in Fig. 5.
}
\label{Fig:PD}
\end{figure}

In Fig. \ref{Fig:PD}, $T_\mathrm{c}$'s are plotted against $\nu _\mathrm{Q3}$ at 77 K for all the A, B and C samples and the different batches shown by open triangles.
Two superconducting phases, labeled by SC-I and SC-II, are separated by an intermediate nonsuperconducting region.
We regard the M phase as a magnetic ordering phase because the occurrence of a weak internal magnetic field is observed at 1.8 K and a magnetic critical slowing down effect at about 5 K in this region (see below).
Open diamonds are magnetic transition temperatures $T_\mathrm{M}$'s estimated from $^{59}$Co NQR spin-lattice relaxation rate 1/$T_1$ measurements.
One should note that all the $T_\mathrm{c}$ vs $\nu _\mathrm{Q3}$ curves trace the universal curve irrespective of the differences in the soft chemical treatments and their duration dependences.
This universal phase diagram indicates that some unique electronic states characterized by $\nu _\mathrm{Q3}$ make the system superconducting or magnetic.

The value of $\nu _\mathrm{Q}$ tells us the Co valence and/or the configuration of oxygen ions surrounding Co ions.
According to Ref. [6], the duration increases oxygen vacancies, the Co valence decreases from about 3.6, and then the superconductivity is abruptly suppressed for the valence of 3.5.
However, the Co valence is estimated to be about 3.4 by redox titrimetry\cite{Takada_oxonium}, and the superconductivity is reported to be changed by an ion exchange of Na$^{+}$ and H$_3$O$^{+}$ while keeping the Co valence  $\sim$ 3.4\cite{Sakurai_HCl}.
If the oxygen deficiency increases with the duration, the $^{59}$Co NQR spectrum must be broadened and show asymmetric tails. 
As in Fig. \ref{Fig:NQRspe}, we observed the monotonic shift of the $^{59}$Co NQR spectrum to a lower frequency side but not any additional broadening.
Thus, it is unlikely that the oxygen content changes with the duration.

The ion exchange between Na$^+$ and H$_3$O$^+$ can change not only the $c$-axis lattice parameter but also oxygen arrangements and/or the thickness of the CoO$_2$ plane \cite{Ihara_NQR}.
The existence of H$_3$O$^+$ ions was suggested by Raman scattering measurements\cite{Takada_oxonium}, and our $^1$H NMR measurements for the change of content of H$_3$O$^+$ ions (not shown here).
Recently, we estimated directly the invariant Co valence of $\sim$ 3.4 for the durations from soft X-ray absorption spectroscopy\cite{XAS}.
Thus, the development of $\nu _\mathrm{Q3}$ results from local change in the oxygen configuration around Co.

From the theoretical aspects, for the thiner CoO$_2$ plane, the $e_g$' energy level of Co ion lifts and makes six hole pockets of Fermi surface near the K points of the Brillouin zone.
A theoretical phase diagram is proposed to reproduce two superconducting phases and the intermediate magnetic phase\cite{Mochizuki_PD}.
One should note that the experimental phase diagram in Fig. \ref{Fig:PD} resembles the theoretical one.
The theoretical spin states of Cooper pairs in the two superconducting phases are suggested to be of spin singlet extended $s$-wave and of spin triplet $p$-wave.

\begin{figure}[t]
\includegraphics[width=8cm]{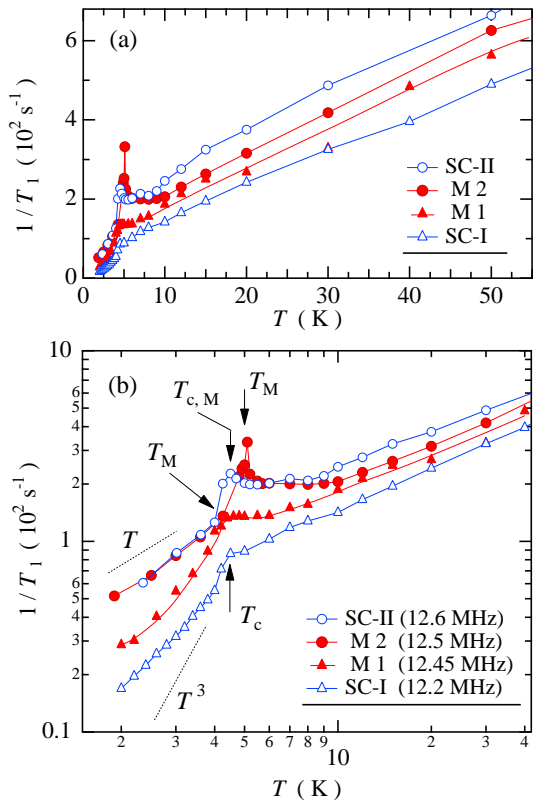}
\caption{
Temperature dependence of $^{59}$Co nuclear spin-lattice relaxation rate 1/$T_1$ of the samples with $\nu _\mathrm{Q3}$ = 12.2 (SC-I), 12.45 (M1), 12.5 (M2) and 12.6 MHz (SC-II) in (a) a linear scale and (b) a log-log plot.
Dotted lines indicate $T$ linear and $T^3$ functions.
}
\label{Fig:rT1_CoNQR}
\end{figure}

Figure \ref{Fig:rT1_CoNQR} shows the temperature dependences of $^{59}$Co nuclear spin-lattice relaxation rates 1/$T_1$'s of the $bi$layer hydrated samples identified by $\nu _\mathrm{Q3}$(77 K) = 12.2 MHz (SC-I), 12.45 MHz (nonsuperconducting M1), 12.5 MHz (nonsuperconducting M2) and 12.6 MHz (SC-II) indicated by arrows in Fig. \ref{Fig:PD}. 
The relaxation time $T_1$ was estimated by fitting a theoretical magnetic relaxation function 
$p(t) = p(0)\{(3/7)\textrm{exp}(-3t/T_1)+(100/77)\textrm{exp}(-10t/T_1)+(3/11)\textrm{exp}(-21t/T_1)\}$
to the experimental recovery curves.
The fitting results were satisfactory for the samples in the paramagnetic states.

In Fig. \ref{Fig:rT1_CoNQR}(a), 1/$T_1$ in the normal state above 10 K shows non-Korringa behavior and systematically increases as $\nu _\mathrm{Q3}$( at 77 K) increases, indicating that the magnetic correlation is enhanced in the SC-II phase more than the SC-I phase.
For the M2 sample, the critical slowing down effect on 1/$T_1$ at $T_\mathrm{M}$ = 5 K evidences the divergence behavior of magnetic correlation length toward a long-range ordering.
To our knowledge, this is the first observation to exclude a charge density wave ordering at $T_\mathrm{M}$.

In Fig. \ref{Fig:rT1_CoNQR}(b), for the M1 and M2 samples, 1/$T_1$ shows a power law behavior of $T^{\alpha}$ (1 $\leq \alpha <$ 3) below $T_\mathrm{M}$, suggesting a Raman process of spin-wave scatterings.
For the SC-I sample below $T_\mathrm{c}$, 1/$T_1$ shows the absence of a Hebel-Slichter peak and a power law of $T^{\alpha}$ (1 $\leq \alpha <$ 3).
For the SC-II sample, 1/$T_1$ shows a peak behavior at $T_\mathrm{c}$, similarly to those for the M1 and M2 samples.
Magnetic ordering seems to coexist with superconducting ordering in the SC-II sample.
Since the systematic shift (SC-II $\to$ M $\to$ SC-I) without any additional broadening in the $^{59}$Co NQR spectrum was observed in Fig. \ref{Fig:NQRspe}(a), the phase separation and the sample inhomogeneity are excluded.
The coexistence of magnetic ordering suggests an unconventional superconductivity in the SC-II phase, because the conventional $s$-wave superconductivity is exclusive with the magnetic ordering.
The Korringa-like behavior in 1/$T_1$ below $T_\mathrm{c, M}$ indicates the residual density of states possibly due to a pair breaking effect of the coexistence of magnetic ordering.
This is similar to ferromagnetic superconductor of UGe$_2$ under a high pressure\cite{Saxena_UGe2, Kotegawa_UGe2}.
The $p$-wave superconductivity might be realized in the SC-II phase.

In conclusion, we found the reentrant behavior of superconductivity of $bi$layer hydrated Na$_x$CoO$_2\cdot y$H$_2$O and the universal $T_\mathrm{c}$ vs $\nu_{Q3}$ phase diagram, where the $^{59}$Co NQR frequency $\nu _\mathrm{Q3}$ characterizes and classifies the various $bi$layer hydrated samples.
Two superconducting phases (SC-I and SC-II) are separated by the in-between magnetic (M) phase in the universal phase diagram.
From $^{59}$Co 1/$T_1$ measurements, we found an evidence of magnetic ordering but not charge ordering in the M phase and a possible indication of a coexistence of superconducting and magnetic ordering.

This work is supported by Grants-in Aid for Scientific Research on Priority Area ``Invention of anomalous quantum materials", from the Ministry of Education, Culture, Sports, Science and Technology of Japan (16076210) and also by Grant-in-Aid for Scientific Research from the Japan Society for Promotion of Science (19350030).
H. Ohta has been supported by Research Fellowships of the Japan Society for the Promotion of Science for Young Scientists.


\begin{thebibliography}{99} 
\bibitem{Takada_Nature} K. Takada $et$ $al$., Nature \textbf{422}, 53 (2003).
\bibitem{Takada_oxonium} K. Takada $et$ $al$., J. Mater. Chem. \textbf{14}, 1448 (2004).
\bibitem{ohta_Duration} H. Ohta $et$ $al$., J. Phys. Soc. Jpn. \textbf{74}, 3147 (2005).
\bibitem{Michioka_NQR} C. Michioka $et$ $al$., J. Phys. Soc. Jpn. \textbf{75}, 063701 (2006).
\bibitem{sakurai_NCO} H. Sakurai $et$ $al$., J. Phys. Soc. Jpn. \textbf{73}, 2081 (2004).
\bibitem{Barnes} P. W. Barnes $et$ $al$., Phys. Rev. B \textbf{72}, 134515 (2005).
\bibitem{Ihara_NQR} Y. Ihara $et$ $al$., J. Phys. Soc. Jpn. \textbf{74}, 867 (2005). 
\bibitem{ohta_icm} H. Ohta $et$ $al$., J. Mag. Mag. Mater. \textbf{310}, e141 (2007).
\bibitem{Ihara_Mthesis} Y. Ihara $et$ $al$., J. Phys. Soc. Jpn. \textbf{75}, 124714 (2006).
\bibitem{Sakurai_HCl} H. Sakurai $et$ $al$., J. Phys. Soc. Jpn. \textbf{74}, 2909 (2005).
\bibitem{XAS} H. Ohta $et$ $al$., to be accepted in PRB.
\bibitem{Mochizuki_PD} M. Mochizuki, and M. Ogata, J. Phys. Soc. Jpn. \textbf{76}, 013704 (2007).
\bibitem{Saxena_UGe2} S. S. Saxena $et$ $al$., Nature \textbf{406}, 587 (2000).
\bibitem{Kotegawa_UGe2} H. Kotegawa $et$ $al$., J. Phys. Soc. Jpn. \textbf{74}, 705 (2005).
\end{thebibliography}
\end{document}